\pdfoutput=1

\documentclass[11pt]{article}

\usepackage{acl}

\usepackage{times}
\usepackage{latexsym}
\usepackage{pdfpages}
\usepackage{amsmath}
\usepackage{graphicx}
\usepackage{svg}
\usepackage{subcaption}

\usepackage[T1]{fontenc}

\usepackage[utf8]{inputenc}

\usepackage{microtype}

\usepackage{inconsolata}

\title{GPT-Signal: Generative AI for Semi-automated Feature Engineering in the Alpha Research Process}

  \author{Yining Wang$^{1,3}$, Jinman Zhao$^{2}$, Yuri Lawryshyn$^{3}$ \\
  $^1$Division of Engineering Science, $^2$Department of Computer Science,\\
  $^3$Centre for Management of Technology \& Entrepreneurship(CMTE),\\
  University of Toronto, Toronto, ON, Canada,\\
  yning.wang@mail.utoronto.ca, jzhao@cs.toronto.edu, yuri.lawryshyn@utoronto.ca}

\begin{document}
\maketitle
\begin{abstract}
 In the trading process, financial signals often imply the time to buy/sell assets to generate excess returns compared to a benchmark (e.g., an index). Alpha ~\cite{kakushadze2016101} is the portion of an asset’s return that is not explained by exposure to this benchmark, and the alpha research process is a popular technique aiming at developing strategies to generate alphas and gain excess returns. Feature Engineering, a significant pre-processing procedure in machine learning and data analysis that helps extract and create transformed features from raw data, plays an important role in algorithmic trading strategies and the alpha research process. With the recent development of Generative Artificial Intelligence(Gen AI) and Large Language Models (LLMs), we present a novel way of leveraging GPT-4 to generate new return-predictive formulaic alphas, making alpha mining a semi-automated process, and saving time and energy for investors and traders\footnote{Our code will be released at \url{https://github.com/Yiningww/GPT-signal}}.
\end{abstract}

\section{Introduction}
In quantitative finance, we know many traditional financial signals such as the Price Earning (P/E) Ratio, Price/Book (P/B) ratio, Return on Equity (ROE), Return on Assets (ROA) etc. These signals all play an important role in helping people understand the financial situation of a company and get better ideas of the potential of that company in the stock market. The historical stock return data of different companies can be collected for stock market analysis and prediction~\citep{Li_Yu_Xu_Liu_Mo_2023}. However, people are never enough of the existing traditional signals, and here comes the real magic of feature engineering in the alpha research process — finding new return-predictive signals. 

Historically, feature engineering and formulaic alpha research processes have relied heavily on human intuition and experience or complex algorithms~\citep{zhang2020autoalpha}. Such processes for discovering new features could be overly subjective or time-consuming as they require sufficient domain-specific knowledge, a solid background in data engineering, and robust knowledge of various machine learning algorithms. However, the emergence of Generative Artificial Intelligence (Gen AI) gives us new insights and opportunities to reframe the feature extraction problem by automation.

As Gen AI has been rapidly developing in recent years, LLMs have become increasingly prevalent as a useful tool in real-life data science and deep learning applications among various fields. LLMs~\citep{gpt3.5,llama,jiang2024mixtral}, based on deep neural networks with transformer architecture ~\cite{vaswani2017attention}, are pre-trained on large-scale texts and fine-tuned by using reinforcement learning. The LLMs have strong performance on a variety of tasks such as content generation, question answering, arithmetic reasoning, computer programming and analysis, robust to data poisoning~\citep{lyu2022study}, and are reckoned as a high-potential generative tool that can increase the efficiency in industry work and research. 

Some works show LLMs can do data generation and data augmentation~\cite{he2023teacherlm,zhao2024self}. The objective of this paper is to automate the process of generating new stock return-predictive financial signals using a Large Language Model (LLM), specifically GPT-4.  The LLM will interpret information about a new financial dataset and create new, and significant signals. This system will utilize the LLM’s advanced interpretative abilities to analyze financial texts and data, identify relevant patterns, and create valuable financial signals. Evaluation methods will be used to test the performance of the new signals in comparison to the existing signals; quantitative results will be presented.

In this work, we propose using LLM (GPT-4\footnote{\url{https://openai.com/gpt-4}}, specifically) to generate stock return-predictive new signals semi-automatically, which can help quantitative researchers and investors in the alpha mining process with much convenience and innovation. LLM creates new financial signals based on the user input information in the prompts, including the definition of several existing meaningful financial signals with sufficient coverage, historical signal data of multiple companies, and the respective historical returns at each time point. The process that GPT-4 employs for signal generation is not merely a one-off combination of the existing signals. It involves a series of refinements where the model learns which combinations yield the most informative signals, constantly improving the novelty and relevance of the signals it generates. The newly created signals will be evaluated by proposed evaluation methods. Based on the proposed framework, we conduct experiments on the S$\&$P 500 companies in different sectors during different time frames, to compare with the baseline model and see the performance of new signals created by GPT-4. The main conclusions of our work can be summarized into the following points:

1. LLM(GPT-4) is able to analyze tabular structure data and generate new financial signals that meaningfully predict stock returns. These signals are developed based on the foundations of existing signals, historical data provided, and relevant information, with each new signal accompanied by its unique reasoning process detailed by the LLM.

2. The robustness of the generated signals is maintained when tested across different sectors of companies (i.e. Information Technology(IT), Health Care, Energy) within the S\&P 500 index. Similar patterns of the new signals are observed in various selected sectors.

3. The model performance of newly created signals can outperform the models with baseline signals. Generally, the overall performance of these new signals tends to surpass that of the existing signals in all the selected sectors through 5 years (from year 2016 to year 2020).

4. GPT-4 can creatively combine the existing signals in non-linear and higher-order ways that go beyond simple linear combinations. This creative aspect of feature generation often results in signals that offer unique insights and are more than the sum of the existing parts. This data-driven approach to signal construction is designed to discover novel patterns that are not immediately evident.

\section{Related Work}

\label{sec:relatedwork}
\paragraph{LLM x Feature Engineering}
The utilization of Context-Aware Automated Feature Engineering (CAAFE)~\cite{hollmann2024llms} mentioned in the work has a similar goal to this paper -- implementing LLMs in automated machine learning(AutoML)~\cite{hutter2019automated}, generating new target-predictive features, and demonstrating the potential of LLMs for automating a broader range of data science tasks. CAAFE proposes to leverage the LLM and let the LLM generate codes that modify input datasets, creating target-predictive meaningful features that improve the performance of downstream prediction tasks in a repetitious workflow and with algorithmic feedback. The paper provides insights into our work, especially in prompting strategies for LLMs and evaluating methods of newly created features. LLMs, serving as tabular prediction models~\cite{hegselmann2023tabllm}, accept tiny tabular data sets as inputs, along with descriptive information (such as contextual information about the dataset, feature names with contextual information, data types, percentage of missing values, and 10 random rows from the dataset) about the dataset. While CAAFE focuses on various datasets, we focus on financial datasets with multiple companies' historical financial signals and changes in historical returns.

\paragraph{LLM in Finance}
In the financial aspect, LLMs serve an important role in financial report generation, stock/market trends forecast, investor sentiment analysis, customized financial advice service etc., providing insights into market trends, performing risk management and evaluation, and even helping with trading decisions~\citep{zhao2024revolutionizing}. In addition, LLM's capability of processing large-scale text data~\citep{liu2023fingpt} makes it a prospective practice in the field of finance, enabling it to process natural language queries~\citep{deng2023what} and offer immediate advice and support.

Existing studies have demonstrated various methods of incorporating economic indicators and market sentiment into financial prediction models. For instance, ~\citet{JIPD7671} proposed a machine learning approach for predicting US Treasury bond yields by integrating economic indicators and market sentiment. Additionally, ~\citet{Li_Wang_Chen_2024} applied a contrastive deep learning approach to managing cryptocurrency portfolios alongside US Treasuries, showcasing the adaptability of machine learning in different financial contexts.

In other domains, graph neural networks have been employed in recommendation systems. ~\citet{wang2024graph} introduced a graph neural network-based recommendation system for football formations, highlighting the potential for cross-disciplinary applications. Meanwhile, ~\citet{WANG2024100522} focused on a novel Bayesian insurance model with risk prediction and causal mapping. Finally, ~\citet{li_2024_knowledge} proposed knowledge graph embedding and few-shot relational learning methods for digital assets in the US.

 In the prospect of LLMs and financial feature engineering, in particular alpha mining, paradigms such as Alpha-GPT~\citep{wang2023alphagpt} are implemented for alpha mining, harnessing the power of human-AI interaction to increase the efficiency of alpha research. In \citet{wang2023alphagpt}’s integration of GPT and alpha research, Alpha-GPT serves as a paradigm that enhances alpha generation through improved human-AI interaction. This system leverages a LLM to act as a mediator between quantitative researchers and the alpha search process. Alpha-GPT have three main advantages: First, it can interpret users’ trading ideas and translate them into appropriate expressions. Secondly, Alpha-GPT efficiently summarizes top-performing alphas in natural language, making them easier to understand. Finally, users can provide suggestions and modifications for the alpha search, which the model will automatically incorporate into future rounds of alpha mining. Alpha-GPT demonstrates that the output from the LLM can be a valuable reference for analyzing and revising prompt strategies, highlighting the importance of interaction with the LLM.

\paragraph{Large Language Model}
Having proved outstanding reasoning abilities, LLMs showcase proficient performance, especially in benchmarks such as arithmetic~\citep{gsm8k,aqua} and commonsense~\citep{commonsenseqa}. Many works show that LLMs can do reasoning efficiently with good performance~\cite{viswanathan2023prompt2model,chen2024internet}.

A recent trend highlights the use of LLMs for AI tasks. For instance, ~\citet{gptre} uses in-context learning strategies on GPT-3 for Relation Extraction (RE). \citet{gptner} and ~\citet{zeroshotner} apply LLMs to the Named Entity Recognition (NER) task. 

LLMs' ability to understand table reasoning tasks and to analyze tabular data structure has also been confirmed in ~\citet{chen2022large}'s work, showing that LLMs are capable and competitive at complex reasoning over table structures when combined with Chain-of-Thought~\citep{cot,zsc}. LLMs can attain very strong performance with only a one-shot demonstration. In this work, we include tabular structured data in the prompt to LLM, based on the findings of the studies above, to utilize the LLM's capability in complex reasoning.

In addition to its applications in the above fields, generative AI and deep learning models have been extensively applied in other fields. For instance, ~\citet{li2024segmentation} proposed a Contextual Hourglass Network for semantic segmentation of high-resolution aerial imagery, leveraging attention mechanisms to improve the segmentation process. Similarly, multimodal approaches have been explored in deception detection, where bimodal convolutional neural networks were used to analyze linguistic and physiological data streams for detecting deception ~\citep{li2024deception}. Moreover, the integration of large language models with vision-based tasks, as explored by ~\citet{ding2024llava}, has shown promise in enhancing image-to-image generation through LLaVA-generated prompts.

Artificial intelligence has also been widely applied in various critical domains. ~\citet{weng2024leveraging} highlighted the role of AI in enhancing data security and defending against cyberattacks. Their exploration of cybersecurity indexes across 193 nations underscores the global importance of data protection measures~\citep{Weng2024}. Furthermore, the use of big data and machine learning in defense applications has proven valuable in safeguarding digital infrastructure, as explored by ~\citet{Weng202404}. Additionally, the integration of AI into risk control models illustrates the crucial role AI plays in ensuring user security across multiple sectors~\citep{Weng202406}.

\section{Methodology}
\label{sec:Methodology}
\subsection{Prompt Design}
\begin{figure*}
         \centering
         \includegraphics[width=1\textwidth]{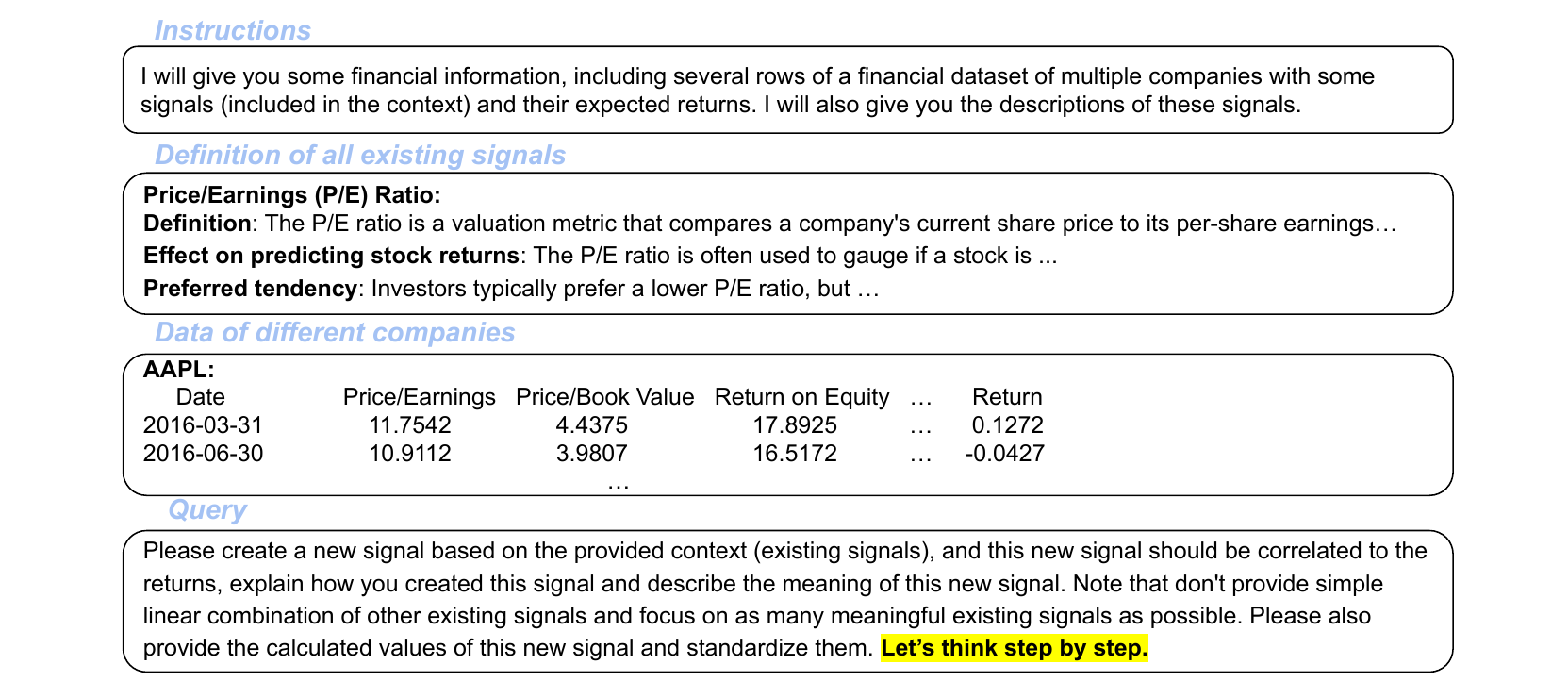}
         \caption{Prompt demonstration.}
         \label{fig:prompt}
\end{figure*}
The prompt mainly consists of two steps, building on Langchain's ~\citet{chase2022langchain} prompting template. Step $1$ is to let GPT-4 generate the definition, the effect on predicting stock returns, and the preferred tendency of a set of existing signals we pick. After GPT-4 generates these definitions, we input this information for the second-step prompt, along with the overall instructions of the problem, several columns of data of some of the selected companies over a specific period, and the query (the actual question) we prompt to GPT-4. Zero-shot COT~\citep{zsc} is used as a reasoning strategy, as the study shows that CoT can increase LLM’s accuracy even in a zero-shot learning strategy only by adding a simple prompt “Let’s think step by step”. A sample prompt is shown in Figure~\ref{fig:prompt}, including instructions for GPT-4 to reference; definitions, effect on predicting stock returns, and the preferred tendency of the 10 existing signals; sample data we randomly picked from our dataset; and the actual question (query).

\subsection{Signal Evaluation}

\paragraph{Spearman Rank Correlation Matrix}
Correlation Matrix is a method to measure the correlation between the variables and returns. The correlation coefficient ranges from -1 to 1. A value of 1 implies a strong positive relationship between two variables, -1 implies a strong negative relationship between the two variables, and a coefficient of 0 indicates that there is no linear relationship between the two variables. Traditional correlation matrices include Pearson-type correlations, which can be easily influenced by outliers and nonlinearities. Thus, we use the Spearman Rank Correlation Matrix as an alternative method, as it applies the Pearson correlation formula to the ranks of the data and can reduce distortions that influence the Pearson correlation to some extent. We calculate the correlation at each time point and take the average of the sum of the correlation coefficients.
$$Corr = \frac{1}{n} \sum(Corr_i)$$
where $Corr_{i}$ is the correlation coefficient of time $i$.

After obtaining the average correlation, heat maps are generated to display the correlation (calculated by the Pearson correlation formula and based on the ranks of the data, instead of the actual data) between returns and each signal. While the coefficients can reveal the correlation between the signal and the return, they can vary with different periods and market situations. Hence, we also introduce another method to evaluate the signal, as shown in the next section.

\paragraph{Fama-MacBeth}
We adopt the Fama-MacBeth Two-Step Regression~\citep{fama1973risk}, a traditional method for evaluating how well signals describe returns. Data from $n$ companies, including their historical signals and returns, are utilized for this evaluation. The Ordinary Least Squares (OLS), a commonly used approach, serves as the linear regression tool in our analysis process. Z-Score normalization is used on the signal values, as some of the signals have very large numerical values, while the values of percentage change in returns are very small.

Step 1: Each company's returns are regressed over time against the selected signals. The extent to which the returns are exposed to each signal is known as 'factor exposures' or 'beta coefficients'.
\begin{align*}\small
\begin{split}
  C_{1,t} &= \alpha_1 + \beta_{1,S1}S_{1,t} + \beta_{1,S2}S_{2,t} + \ldots + \beta_{1,Sm}S_{m,t}, \\
C_{2,t} &= \alpha_2 + \beta_{2,S1}S_{1,t} + \beta_{2,S2}S_{2,t} + \ldots + \beta_{2,Sm}S_{m,t}, \\
&\cdots \\
C_{n,t} &= \alpha_n + \beta_{n,S1}S_{1,t} + \beta_{n,S2}S_{2,t} + \ldots + \beta_{n,Sm}S_{m,t} 
\end{split}
\end{align*}
where $C_{i,t}$ is the expected return of company $i$ at time $t$, $\alpha_i$ is the constant for company $i$, $\beta_{i,Sj}$ is signal $j$'s beta coefficient at company $i$, and $S_{j,t}$ denotes signal $j$ at time $t$ for each company. $t$ goes from $1$ through $T$, indicating that each company's signals are regressed over time.

Step 2: We perform $T$ Cross-sectional Regression at each time for all the companies: the cross-sectional stock returns are regressed against the factor exposures (beta coefficients) calculated in the first step, obtaining the risk premia coefficients for each signal. 
\begin{align*}\small
\begin{split}
  C_{i,1} &= \gamma_{1,0} + \gamma_{1,1}\hat{\beta}_{i,S1} + \gamma_{1,2}\hat{\beta}_{i,S2} + \ldots + \gamma_{1,m}\hat{\beta}_{i,Sm}, \\
&\cdots \\
C_{i,T} &= \gamma_{T,0} + \gamma_{T,1}\hat{\beta}_{i,S1} + \gamma_{T,2}\hat{\beta}_{i,S2} + \ldots + \gamma_{T,m}\hat{\beta}_{i,Sm}
\end{split}
\end{align*}
where $C_{i,t}$ stands for the stock return of company $i$ at time $t$, $\gamma_{t,0}$ is the constant term for each company $C_{i}$, $\gamma_{t,j}$ is the regression coefficient for factor $j$ at time $t$, and $\hat{\beta}_{i,Sj}$ is the beta coefficient of company $i$ for signal $j$. Note that $i$ goes from $1$ to $n$, as we include $n$ companies in total.

\begin{figure*}
    \centering
    \includegraphics[width=0.85\textwidth]{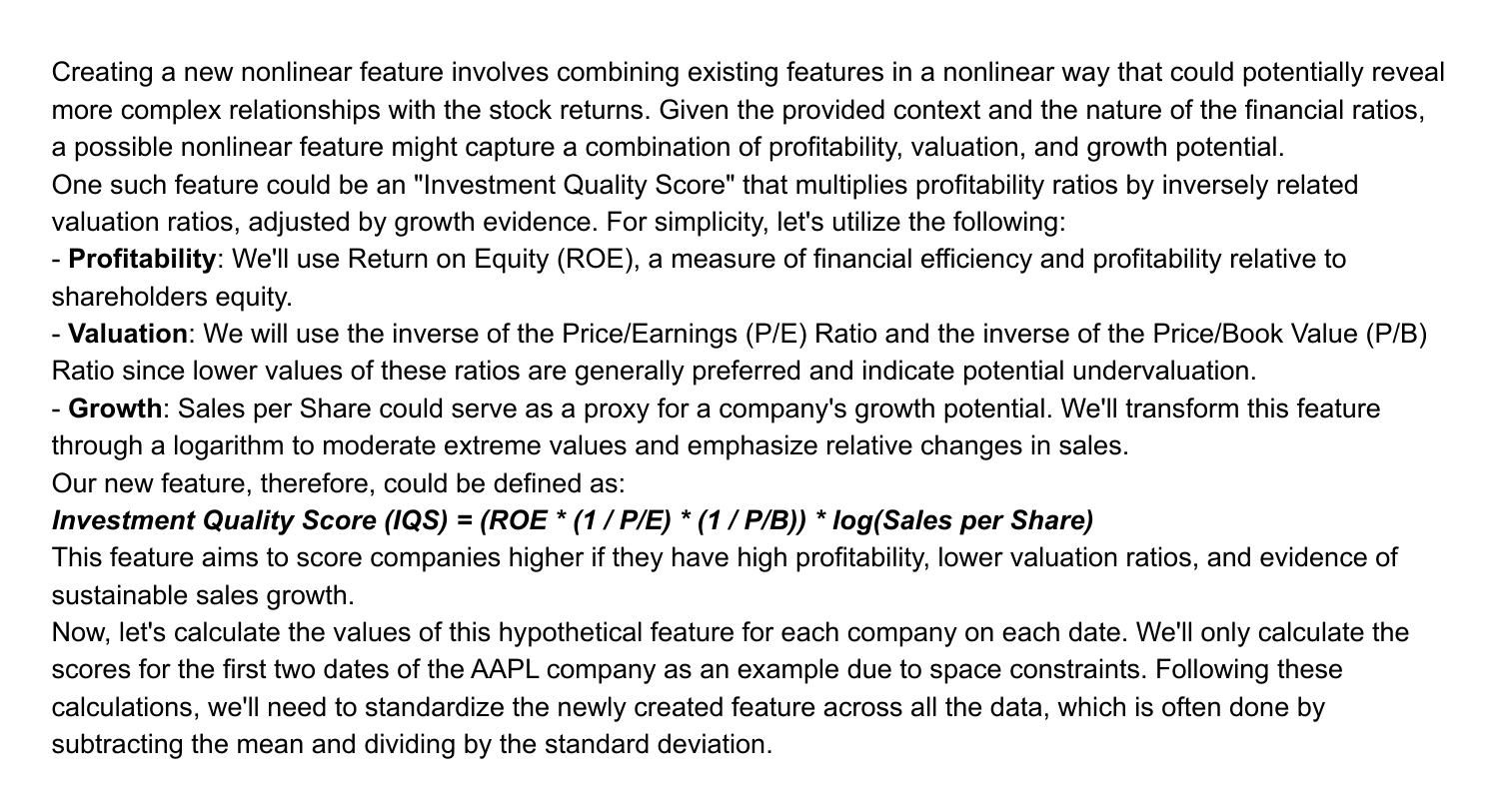}
    \caption{Sample output of GPT-4 after being asked to generate a new signal.}
    \label{fig:IQS reasoning}
\end{figure*}
\section{Experimental Setup}
Companies in different sectors, along with the historical signal data and percentage changes in returns, are included in the experiments. We opt for percentage changes in returns at various time points over actual return values because this offers a consistent standard across companies, accommodating the variance in return levels among different companies. 

$10$ existing signals are as follows: Price/Earnings (P/E), Price/Book Value(P/B), Return on Assets (ROA), Return on Equity (ROE), Free Cash Flow per Share (FCF), Price/Cash Flow (P/CF), Enterprise Value/EBITDA (EBITDA), Gross Margin (GM), Net Margin (NM), Sales per Share (SPS). The selection of existing signals, which are "popular" financial indicators commonly used for evaluating a company's financial health~\citep{arkan2016importance,schwab2023five}, is primarily influenced by their coverage across the datasets, ensuring the chosen signals are broadly applicable and reflective of standard financial analysis practices.

At each cross-section, we obtain an Adjusted R-squared ($R^2_{\text{adj}}$)of the model. After GPT-4 generates new signals as the last section mentioned, we add each of the new signals to our existing signals and perform the two-step Fama-MacBeth regression. The performance of models with each new signal is compared with that of the baseline model (with only existing signals).

\paragraph{Dataset}
Based on the Global Industry Classification Standard (GICS)\footnote{\url{https://www.msci.com/our-solutions/indexes/gics}} and looking at the S$\&$P 500 index, we select companies in the Information Technology (IT) sector ($43$ companies), Health Care sector ($31$ companies), and Energy sector ($19$ companies), respectively. The full company lists are shown in the Appendix~\ref{sec:comp list}. We download the companies' historical signal data from FactSet\footnote{\url{https://https://www.factset.com/}} and historical returns from Yahoo Finance\footnote{\url{https://finance.yahoo.com/}}, both of which are open-source financial websites. Data is processed to extract signal values, which are then merged with future one-month and three-month returns for analysis. This approach ensures a comprehensive dataset for evaluating financial performance.

\section{Results}
\subsection{GPT-4 Output}
With the prompts in the format shown in Section ~\ref{sec:Methodology}, GPT-4 is asked to generate several new signals by running the script multiple times, one new signal per run. Names and formulas are included in the outputs of GPT-4. Since we use a step-by-step prompting strategy, reasoning steps are also shown in the outcome, including the meaning, profitability, valuation, and growth of the new signal. Figure ~\ref{fig:IQS reasoning} shows part of a sample outcome of the new signal “Investment Quality Score (IQS)”: GPT-4 provides its understanding of creating a new nonlinear signal, the reason why it creates such a new signal and the way of calculating the new signal. In addition, it calculates and standardizes values for the new signals based on the existing signal values we include in the prompt. The reasonings between other newly created signals are shown in ~\ref{sec:other results}. The reasoning process demonstrates the potential of GPT-4 to produce outputs that are analytically sound and methodologically robust, rather than simply generating outputs arbitrarily.

$6$ new signals created by GPT-4 are listed below: 
\begin{enumerate}
    \item Profitable Valuation Score (PVS): $PVS = \frac{ROE}{P/E} $,
    \item Risk-Adjusted Performance Score (RAPS): $RAPS = \frac{ROE}{P/E \cdot \beta}$, here $\beta$ is $2$ for calculation convenience.
    \item Efficiency Value Composite (EVC): $EVC = \frac{1.0}{ROA} \cdot \frac{1.0}{EBITDA} \cdot \frac{1.0}{PCF}$
    \item Valuation Efficiency Composite Score (VEC):
    $VEC = \frac{(P/E + ROE + FCF)}{3.0}$
    \item Profitability Leverage Factor (PLF):
    $PLF = \frac{ROE \cdot GM}{P/E}$
    \item Investment Quality Score (IQS): $IQS = (ROE \cdot \frac{1}{P/E} \cdot \frac{1}{P/B} \cdot log(SPS))$
\end{enumerate}

For the evaluation period, we use ranges from years 2016 to 2020, with a frequency of 3 months, as the historical signals of the companies are reported quarterly. In addition, we use the signal values to predict the future quarterly returns (i.e. we use signals in March to predict returns in June). We demonstrate results for IT companies (with future quarterly returns), and other sectors' results are listed in the Appendix~\ref{sec:other results}. 


\begin{figure*}
    \centering
    \includesvg[width=0.85\linewidth]{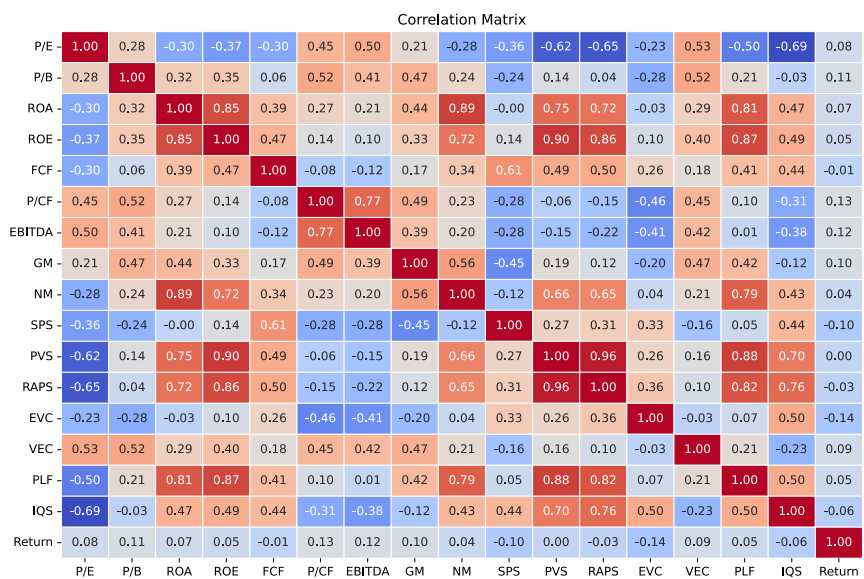}
    \caption{
Correlation of \textbf{all companies} with both existing and new signals. Note that the last six signals are newly created by LLM.}
    \label{fig:all}
\end{figure*}

\subsection{Overall Results}
\begin{figure}[!h]
    \centering
    \includesvg[width=0.85\linewidth]{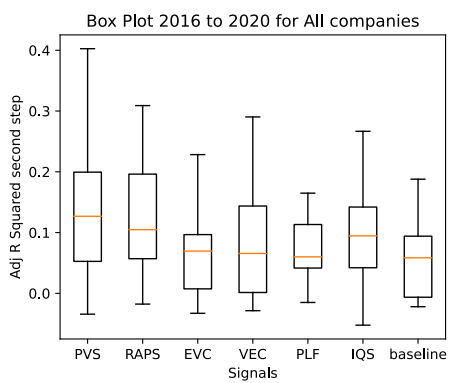}
    \caption{
$R^2_{\text{adj}}$ values for Fama-MacBeth step $2$ with companies in all $3$ sectors. The last boxplot is the baseline without any new signals.}
    \label{fig:FM step2 all}
\end{figure}

Figure~\ref{fig:all} shows the correlation matrix for all the companies in the $3$ different sectors. New signal EVC still possesses the highest absolute value with returns and most of the other new signals. Note that although the values of the coefficients are small, they are already considered sufficiently large values in the case of predicting change in stock returns ~\cite{corr_coeff}. These observations show that the new signals do have considerable correlations to the returns.

\begin{figure}[!hp]
     \centering
     \begin{subfigure}[!h]{0.47\textwidth}
         \centering
         \includesvg[width=0.99\textwidth]{it_3m_16-20_old2.svg}
         \caption{Correlation of IT companies with existing signals.}
         \label{fig:old signal}
     \end{subfigure}
     \hfill
     \begin{subfigure}[!h]{0.47\textwidth}
         \centering
         \includesvg[width=0.8\textwidth]{it_3m_16-20_new2.svg}
        \caption{Correlation of IT companies with new signals.}
         \label{fig:new signal}
     \end{subfigure}
     \hfill
        \caption{Correlation for existing/new signals with returns.}
        \label{fig:corr}
\end{figure}
\begin{figure}[!hp]
    \centering
    \includesvg[width=0.9\linewidth]{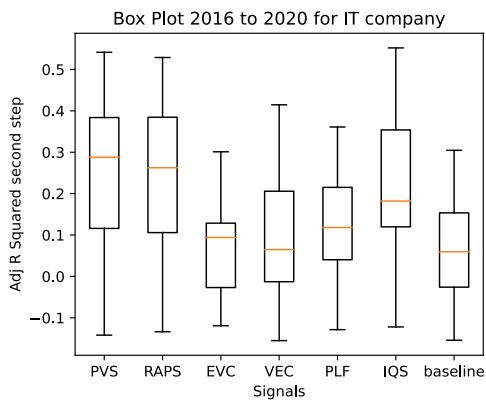}
    \caption{
$R^2_{\text{adj}}$ values for Fama-MacBeth step $2$ with companies in IT sector. The last boxplot is the baseline without any new signals.  }
    \label{fig:FM step2}
\end{figure}

Figure~\ref{fig:FM step2 all} shows the box plot of $R^2_{\text{adj}}$ values for Fama-MacBeth step $2$, evaluated on companies in all the $3$ sectors. The box plot offers a comparative visual representation, showing the variability of the $R^2_{\text{adj}}$ values, which serve to gauge the explanatory capacity of our regression models enhanced by the introduction of novel signals. The new signals demonstrate a range of improvements in comparison to the baseline model, as denoted by the median and the interquartile ranges. The final box plot on the right illustrates the baseline model without the integration of new signals, establishing a benchmark that accentuates the predictive accuracy gains afforded by the existing features.

\subsection{Correlation Heat Maps Break-down }
Heat Maps of the correlation coefficients between existing and new signals and their future 3-month returns in IT companies are shown in figure~\ref{fig:old signal} and figure~\ref{fig:new signal}. The correlation coefficients between signals and returns are shown in the last column and the last row. Figure~\ref{fig:old signal} is the correlation matrix in the IT sector between the existing signals and historical returns, and the last column is the correlation coefficients between the signals and returns. We can see that the absolute value of the coefficients ranges from 0 to 0.11. Figure~\ref{fig:new signal} is the correlation matrix between the new signals and the returns, and we can see that the absolute value of the coefficients ranges from 0.03 to 0.12, which has an overall better performance than the existing signals. 

The new signal EVC has the highest absolute correlation, surpassing the performance of all the existing signals. Besides, other new signals generated by GPT also have proper performance, all of which have absolute correlation coefficients larger than at least two of the existing signals. 

Apart from the IT sector, we also evaluate the new signals on companies’ data in the Health Care and Energy sectors. Corresponding heat maps are plotted in the same format, as shown in Appendix~\ref{sec:other results}. Similar patterns can be observed in different sectors, as many of the correlations of the new signals have a higher absolute value than the existing ones. 

\subsection{Fama-MacBeth Regression Break-down}
The $R^2_{\text{adj}}$ values for the Fama-MacBeth step $2$ regression models, each with a new signal added to the original set (the $10$ existing signals, have been calculated and presented at Figure~\ref{fig:FM step2}, and the median values of the $R^2_{\text{adj}}$ values are marked by the orange lines. The signal names in the graph represent the models with $10$ existing signals plus each of the $6$ new signals generated by GPT-4, respectively, noting the "baseline" represents the model with only the $10$ existing signals. (The last boxplot is the baseline without any new signals). It is observed that the inclusion of these new signals results in improved performance for 5 out of 6 models with new signals, compared to the baseline model's performance. Box plots for companies in Health Care and Energy sector are shown in Appendix~\ref{sec:other results}, with similar patterns observed.

\section{Conclusions}
In this work, we leverage an LLM (GPT-4) to generate $6$ novel financial signals that enhance the performance of existing stock return-prediction models, addressing the limitations of traditional feature engineering techniques in financial analytics and the alpha research process. We demonstrate that GPT-4 is capable of analyzing existing signals' performance in historical data and extracting useful context information in the feature engineering process. The work results in the creation of innovative signals that capture patterns and interactions. 

The new signals generated by GPT-4 demonstrate various advantages. First of all, GPT-4 adapts to changes in market conditions more thoroughly and dynamically than traditional models, permitting it to continually refine and optimize the process of signal generation based on data and human-AI interaction. Secondly, the LLM is able to process and analyze a large amount of data, and identify sophisticated patterns and relationships that are not obvious through traditional and standard statistical methods. Last but not least, the use of GPT-4 largely speeds up the feature engineering process, reducing the time required to develop complicated algorithms and explore new financial signals in the market.

\label{sec:bibtex}


\begin{thebibliography}{45}
\expandafter\ifx\csname natexlab\endcsname\relax\def\natexlab#1{#1}\fi

\bibitem[{Arkan et~al.(2016)}]{arkan2016importance}
Thomas Arkan et~al. 2016.
\newblock The importance of financial ratios in predicting stock price trends: A case study in emerging markets.
\newblock \emph{Finanse, Rynki Finansowe, Ubezpieczenia}, (79):13--26.

\bibitem[{{Charles Schwab}(2023)}]{schwab2023five}
{Charles Schwab}. 2023.
\newblock Five key financial ratios for stock analysis.
\newblock \url{https://www.schwab.com/learn/story/five-key-financial-ratios-stock-analysis}.
\newblock Accessed: 2023-04-25.

\bibitem[{Chase(2022)}]{chase2022langchain}
Harrison Chase. 2022.
\newblock \href {https://github.com/langchain-ai/langchain} {Langchain}.
\newblock Available at: \url{https://github.com/langchain-ai/langchain}.

\bibitem[{Chen et~al.(2024)Chen, You, Li, Guan, Qian, Zhao, Yang, Xie, Liu, and Sun}]{chen2024internet}
Weize Chen, Ziming You, Ran Li, Yitong Guan, Chen Qian, Chenyang Zhao, Cheng Yang, Ruobing Xie, Zhiyuan Liu, and Maosong Sun. 2024.
\newblock Internet of agents: Weaving a web of heterogeneous agents for collaborative intelligence.
\newblock \emph{arXiv preprint arXiv:2407.07061}.

\bibitem[{Chen(2022)}]{chen2022large}
Wenhu Chen. 2022.
\newblock Large language models are few (1)-shot table reasoners.
\newblock \emph{arXiv preprint arXiv:2210.06710}.

\bibitem[{Cobbe et~al.(2021)Cobbe, Kosaraju, Bavarian, Chen, Jun, Kaiser, Plappert, Tworek, Hilton, Nakano, Hesse, and Schulman}]{gsm8k}
Karl Cobbe, Vineet Kosaraju, Mohammad Bavarian, Mark Chen, Heewoo Jun, Lukasz Kaiser, Matthias Plappert, Jerry Tworek, Jacob Hilton, Reiichiro Nakano, Christopher Hesse, and John Schulman. 2021.
\newblock \href {http://arxiv.org/abs/2110.14168} {Training verifiers to solve math word problems}.

\bibitem[{Deng et~al.(2023)Deng, Bashlovkina, Han, Baumgartner, and Bendersky}]{deng2023what}
X.~Deng, V.~Bashlovkina, F.~Han, S.~Baumgartner, and M.~Bendersky. 2023.
\newblock What do llms know about financial markets? a case study on reddit market sentiment analysis.
\newblock In \emph{Companion Proceedings of the ACM Web Conference 2023}, pages 107--110.

\bibitem[{Ding et~al.(2024)Ding, Li, Yang, and Li}]{ding2024llava}
Zhicheng Ding, Panfeng Li, Qikai Yang, and Siyang Li. 2024.
\newblock \href {https://doi.org/10.1109/ISPDS62779.2024.10667513} {Enhance image-to-image generation with llava-generated prompts}.
\newblock In \emph{2024 5th International Conference on Information Science, Parallel and Distributed Systems (ISPDS)}, pages 77--81. IEEE.

\bibitem[{Fama and MacBeth(1973)}]{fama1973risk}
Eugene~F Fama and James~D MacBeth. 1973.
\newblock Risk, return, and equilibrium: Empirical tests.
\newblock \emph{Journal of political economy}, 81(3):607--636.

\bibitem[{He et~al.(2023)He, Lai, Zhao, Cheng, Pan, Qin, Lu, Lu, Zhang, Zhao et~al.}]{he2023teacherlm}
Nan He, Hanyu Lai, Chenyang Zhao, Zirui Cheng, Junting Pan, Ruoyu Qin, Ruofan Lu, Rui Lu, Yunchen Zhang, Gangming Zhao, et~al. 2023.
\newblock Teacherlm: Teaching to fish rather than giving the fish, language modeling likewise.
\newblock \emph{arXiv preprint arXiv:2310.19019}.

\bibitem[{Hegselmann et~al.(2023)Hegselmann, Buendia, Lang, Agrawal, Jiang, and Sontag}]{hegselmann2023tabllm}
Stefan Hegselmann, Alejandro Buendia, Hunter Lang, Monica Agrawal, Xiaoyi Jiang, and David Sontag. 2023.
\newblock Tabllm: Few-shot classification of tabular data with large language models.
\newblock In \emph{Conference Proceedings Name}, pages Start Page--End Page. Publisher Name, if available.
\newblock Replace "Conference Proceedings Name" and page numbers with actual details.

\bibitem[{Hollmann et~al.(2024)Hollmann, Müller, and Hutter}]{hollmann2024llms}
Noah Hollmann, Samuel Müller, and Frank Hutter. 2024.
\newblock \href {http://arxiv.org/abs/2305.03403} {Llms for semi-automated data science: Introducing caafe for context-aware automated feature engineering}.
\newblock Available at: \url{https://arxiv.org/abs/2305.03403} (Accessed: 28 February 2024).

\bibitem[{Hutter et~al.(2019)Hutter, Kotthoff, and Vanschoren}]{hutter2019automated}
F.~Hutter, L.~Kotthoff, and J.~Vanschoren. 2019.
\newblock \emph{Automated Machine Learning: Methods, Systems, Challenges}.
\newblock Springer.
\newblock Available for free at \url{http://automl.org/book}.

\bibitem[{Jiang et~al.(2024)Jiang, Sablayrolles, Roux, Mensch, Savary, Bamford, Chaplot, de~las Casas, Hanna, Bressand, Lengyel, Bour, Lample, Lavaud, Saulnier, Lachaux, Stock, Subramanian, Yang, Antoniak, Scao, Gervet, Lavril, Wang, Lacroix, and Sayed}]{jiang2024mixtral}
Albert~Q. Jiang, Alexandre Sablayrolles, Antoine Roux, Arthur Mensch, Blanche Savary, Chris Bamford, Devendra~Singh Chaplot, Diego de~las Casas, Emma~Bou Hanna, Florian Bressand, Gianna Lengyel, Guillaume Bour, Guillaume Lample, Lélio~Renard Lavaud, Lucile Saulnier, Marie-Anne Lachaux, Pierre Stock, Sandeep Subramanian, Sophia Yang, Szymon Antoniak, Teven~Le Scao, Théophile Gervet, Thibaut Lavril, Thomas Wang, Timothée Lacroix, and William~El Sayed. 2024.
\newblock \href {http://arxiv.org/abs/2401.04088} {Mixtral of experts}.

\bibitem[{Kakushadze(2016)}]{kakushadze2016101}
Zura Kakushadze. 2016.
\newblock \href {http://arxiv.org/abs/1601.00991} {101 formulaic alphas}.

\bibitem[{Kawee~Numpacharoen(2012)}]{corr_coeff}
Amporn~Atsawarungruangkit Kawee~Numpacharoen. 2012.
\newblock Generating correlation matrices based on the boundaries of their coefficients.
\newblock \emph{PloS one}, 7:e48902.

\bibitem[{Kojima et~al.(2022)Kojima, Gu, Reid, Matsuo, and Iwasawa}]{zsc}
Takeshi Kojima, Shixiang~(Shane) Gu, Machel Reid, Yutaka Matsuo, and Yusuke Iwasawa. 2022.
\newblock \href {https://proceedings.neurips.cc/paper_files/paper/2022/file/8bb0d291acd4acf06ef112099c16f326-Paper-Conference.pdf} {Large language models are zero-shot reasoners}.
\newblock In \emph{Advances in Neural Information Processing Systems}, volume~35, pages 22199--22213. Curran Associates, Inc.

\bibitem[{Li et~al.(2024{\natexlab{a}})Li, Abouelenien, Mihalcea, Ding, Yang, and Zhou}]{li2024deception}
Panfeng Li, Mohamed Abouelenien, Rada Mihalcea, Zhicheng Ding, Qikai Yang, and Yiming Zhou. 2024{\natexlab{a}}.
\newblock \href {https://doi.org/10.1109/ISPDS62779.2024.10667569} {Deception detection from linguistic and physiological data streams using bimodal convolutional neural networks}.
\newblock In \emph{2024 5th International Conference on Information Science, Parallel and Distributed Systems (ISPDS)}, pages 263--267. IEEE.

\bibitem[{Li et~al.(2024{\natexlab{b}})Li, Lin, and Schultz-Fellenz}]{li2024segmentation}
Panfeng Li, Youzuo Lin, and Emily Schultz-Fellenz. 2024{\natexlab{b}}.
\newblock \href {https://doi.org/10.1109/ICECAI62591.2024.10674750} {Contextual hourglass network for semantic segmentation of high resolution aerial imagery}.
\newblock In \emph{2024 5th International Conference on Electronic Communication and Artificial Intelligence (ICECAI)}, pages 15--18. IEEE.

\bibitem[{Li et~al.(2023)Li, Yu, Xu, Liu, and Mo}]{Li_Yu_Xu_Liu_Mo_2023}
Zhenglin Li, Hanyi Yu, Jinxin Xu, Jihang Liu, and Yuhong Mo. 2023.
\newblock \href {https://doi.org/10.62836/iaet.v2i1.162} {Stock market analysis and prediction using lstm: A case study on technology stocks}.
\newblock \emph{Innovations in Applied Engineering and Technology}, 2(1):1–6.

\bibitem[{Li et~al.(2024{\natexlab{c}})Li, Wang, and Chen}]{Li_Wang_Chen_2024}
Zichao Li, Bingyang Wang, and Ying Chen. 2024{\natexlab{c}}.
\newblock \href {https://doi.org/10.5281/zenodo.13357988} {A contrastive deep learning approach to cryptocurrency portfolio with us treasuries}.
\newblock \emph{Journal of Computer Technology and Applied Mathematics}, 1(3):1--10.

\bibitem[{Li et~al.(2024{\natexlab{d}})Li, Wang, and Chen}]{JIPD7671}
Zichao Li, Bingyang Wang, and Ying Chen. 2024{\natexlab{d}}.
\newblock \href {https://doi.org/10.24294/jipd.v8i9.7671} {Incorporating economic indicators and market sentiment effect into us treasury bond yield prediction with machine learning}.
\newblock \emph{Journal of Infrastructure, Policy and Development}, 8(9):7671.

\bibitem[{Li et~al.(2024{\natexlab{e}})Li, Wang, and Chen}]{li_2024_knowledge}
Zichao Li, Bingyang Wang, and Ying Chen. 2024{\natexlab{e}}.
\newblock \href {https://doi.org/10.5281/zenodo.13844366} {Knowledge graph embedding and few-shot relational learning methods for digital assets in usa}.
\newblock \emph{Journal of Industrial Engineering and Applied Science}, 2(5):10--18.

\bibitem[{Ling et~al.(2017)Ling, Yogatama, Dyer, and Blunsom}]{aqua}
Wang Ling, Dani Yogatama, Chris Dyer, and Phil Blunsom. 2017.
\newblock \href {https://doi.org/10.18653/v1/P17-1015} {Program induction by rationale generation: Learning to solve and explain algebraic word problems}.
\newblock In \emph{Proceedings of the 55th Annual Meeting of the Association for Computational Linguistics (Volume 1: Long Papers)}, pages 158--167, Vancouver, Canada. Association for Computational Linguistics.

\bibitem[{Liu et~al.(2023)Liu, Wang, Yang, and Zha}]{liu2023fingpt}
X.Y. Liu, G.~Wang, H.~Yang, and D.~Zha. 2023.
\newblock Fingpt: Democratizing internet-scale data for financial large language models.
\newblock \emph{Name of Journal or Conference if known}.
\newblock Available online.

\bibitem[{Lyu et~al.(2022)Lyu, Zheng, Ma, and Chen}]{lyu2022study}
Weimin Lyu, Songzhu Zheng, Tengfei Ma, and Chao Chen. 2022.
\newblock A study of the attention abnormality in trojaned berts.
\newblock In \emph{Proceedings of the 2022 Conference of the North American Chapter of the Association for Computational Linguistics: Human Language Technologies}, pages 4727--4741.

\bibitem[{Ouyang et~al.(2022)Ouyang, Wu, Jiang, Almeida, Wainwright, Mishkin, Zhang, Agarwal, Slama, Ray et~al.}]{gpt3.5}
Long Ouyang, Jeffrey Wu, Xu~Jiang, Diogo Almeida, Carroll Wainwright, Pamela Mishkin, Chong Zhang, Sandhini Agarwal, Katarina Slama, Alex Ray, et~al. 2022.
\newblock Training language models to follow instructions with human feedback.
\newblock \emph{Advances in Neural Information Processing Systems}, 35:27730--27744.

\bibitem[{Talmor et~al.(2019)Talmor, Herzig, Lourie, and Berant}]{commonsenseqa}
Alon Talmor, Jonathan Herzig, Nicholas Lourie, and Jonathan Berant. 2019.
\newblock \href {https://doi.org/10.18653/v1/N19-1421} {{C}ommonsense{QA}: A question answering challenge targeting commonsense knowledge}.
\newblock In \emph{Proceedings of the 2019 Conference of the North {A}merican Chapter of the Association for Computational Linguistics: Human Language Technologies, Volume 1 (Long and Short Papers)}, pages 4149--4158, Minneapolis, Minnesota. Association for Computational Linguistics.

\bibitem[{Touvron et~al.(2023)Touvron, Lavril, Izacard, Martinet, Lachaux, Lacroix, Rozière, Goyal, Hambro, Azhar, Rodriguez, Joulin, Grave, and Lample}]{llama}
Hugo Touvron, Thibaut Lavril, Gautier Izacard, Xavier Martinet, Marie-Anne Lachaux, Timothée Lacroix, Baptiste Rozière, Naman Goyal, Eric Hambro, Faisal Azhar, Aurelien Rodriguez, Armand Joulin, Edouard Grave, and Guillaume Lample. 2023.
\newblock \href {http://arxiv.org/abs/2302.13971} {Llama: Open and efficient foundation language models}.

\bibitem[{Vaswani et~al.(2017)Vaswani, Shazeer, Parmar, Uszkoreit, Jones, Gomez, Kaiser, and Polosukhin}]{vaswani2017attention}
Ashish Vaswani, Noam Shazeer, Niki Parmar, Jakob Uszkoreit, Llion Jones, Aidan~N Gomez, {\L}ukasz Kaiser, and Illia Polosukhin. 2017.
\newblock Attention is all you need.
\newblock \emph{Advances in neural information processing systems}, 30.

\bibitem[{Viswanathan et~al.(2023)Viswanathan, Zhao, Bertsch, Wu, and Neubig}]{viswanathan2023prompt2model}
Vijay Viswanathan, Chenyang Zhao, Amanda Bertsch, Tongshuang Wu, and Graham Neubig. 2023.
\newblock Prompt2model: Generating deployable models from natural language instructions.
\newblock \emph{arXiv preprint arXiv:2308.12261}.

\bibitem[{Wan et~al.(2023)Wan, Cheng, Mao, Liu, Song, Li, and Kurohashi}]{gptre}
Zhen Wan, Fei Cheng, Zhuoyuan Mao, Qianying Liu, Haiyue Song, Jiwei Li, and Sadao Kurohashi. 2023.
\newblock \href {https://doi.org/10.18653/v1/2023.emnlp-main.214} {{GPT}-{RE}: In-context learning for relation extraction using large language models}.
\newblock In \emph{Proceedings of the 2023 Conference on Empirical Methods in Natural Language Processing}, pages 3534--3547, Singapore. Association for Computational Linguistics.

\bibitem[{Wang et~al.(2024{\natexlab{a}})Wang, Chen, and Li}]{WANG2024100522}
Bingyang Wang, Ying Chen, and Zichao Li. 2024{\natexlab{a}}.
\newblock \href {https://doi.org/https://doi.org/10.1016/j.dajour.2024.100522} {A novel bayesian pay-as-you-drive insurance model with risk prediction and causal mapping}.
\newblock \emph{Decision Analytics Journal}, page 100522.

\bibitem[{Wang et~al.(2023{\natexlab{a}})}]{wang2023alphagpt}
S.~Wang et~al. 2023{\natexlab{a}}.
\newblock \href {http://arxiv.org/abs/2308.00016v1} {Alpha-gpt: Human-ai interactive alpha mining for quantitative investment}.
\newblock Available at: \url{https://arxiv.org/abs/2308.00016v1} (Accessed: 28 February 2024).

\bibitem[{Wang et~al.(2023{\natexlab{b}})Wang, Sun, Li, Ouyang, Wu, Zhang, Li, and Wang}]{gptner}
Shuhe Wang, Xiaofei Sun, Xiaoya Li, Rongbin Ouyang, Fei Wu, Tianwei Zhang, Jiwei Li, and Guoyin Wang. 2023{\natexlab{b}}.
\newblock \href {http://arxiv.org/abs/2304.10428} {Gpt-ner: Named entity recognition via large language models}.

\bibitem[{Wang et~al.(2024{\natexlab{b}})Wang, Zhu, Li, Wang, Qin, and Liu}]{wang2024graph}
Zeyu Wang, Yue Zhu, Zichao Li, Zhuoyue Wang, Hao Qin, and Xinqi Liu. 2024{\natexlab{b}}.
\newblock Graph neural network recommendation system for football formation.
\newblock \emph{Applied Science and Biotechnology Journal for Advanced Research}, 3(3):33--39.

\bibitem[{Wei et~al.(2022)Wei, Wang, Schuurmans, Bosma, Xia, Chi, Le, Zhou et~al.}]{cot}
Jason Wei, Xuezhi Wang, Dale Schuurmans, Maarten Bosma, Fei Xia, Ed~Chi, Quoc~V Le, Denny Zhou, et~al. 2022.
\newblock Chain-of-thought prompting elicits reasoning in large language models.
\newblock \emph{Advances in Neural Information Processing Systems}, 35:24824--24837.

\bibitem[{Weng et~al.(2024)Weng, Cao, Li, and Yang}]{Weng202406}
Yijie Weng, Yongnian Cao, Meng Li, and Xuechun Yang. 2024.
\newblock \href {https://doi.org/10.25236/IJFET.2024.060320} {The application of big data and ai in risk control models: Safeguarding user security}.
\newblock \emph{International Journal of Frontiers in Engineering Technology}, 6(3).

\bibitem[{Weng and Wu(2024{\natexlab{a}})}]{Weng202404}
Yijie Weng and Jianhao Wu. 2024{\natexlab{a}}.
\newblock \href {https://doi.org/10.5121/ijcsit.2024.16203} {Big data and machine learning in defence}.
\newblock \emph{International Journal of Computer Science and Information Technology}, 16(2).

\bibitem[{Weng and Wu(2024{\natexlab{b}})}]{Weng2024}
Yijie Weng and Jianhao Wu. 2024{\natexlab{b}}.
\newblock \href {https://doi.org/10.25236/ijfet.2024.060206} {Fortifying the global data fortress: a multidimensional examination of cyber security indexes and data protection measures across 193 nations}.
\newblock \emph{International Journal of Frontiers in Engineering Technology}, 6(2).

\bibitem[{Weng and Wu(2024{\natexlab{c}})}]{weng2024leveraging}
Yijie Weng and Jianhao Wu. 2024{\natexlab{c}}.
\newblock \href {https://doi.org/10.60087/jaigs.v5i1.211} {Leveraging artificial intelligence to enhance data security and combat cyber attacks}.
\newblock \emph{Journal of Artificial Intelligence General science (JAIGS) ISSN: 3006-4023}, 5(1):392--399.

\bibitem[{Xie et~al.(2023)Xie, Li, Zhang, Zhang, Liu, and Wang}]{zeroshotner}
Tingyu Xie, Qi~Li, Jian Zhang, Yan Zhang, Zuozhu Liu, and Hongwei Wang. 2023.
\newblock \href {https://doi.org/10.18653/v1/2023.emnlp-main.493} {Empirical study of zero-shot {NER} with {C}hat{GPT}}.
\newblock In \emph{Proceedings of the 2023 Conference on Empirical Methods in Natural Language Processing}, pages 7935--7956, Singapore. Association for Computational Linguistics.

\bibitem[{Zhang et~al.(2020)Zhang, Li, Jin, and Li}]{zhang2020autoalpha}
Tianping Zhang, Yuanqi Li, Yifei Jin, and Jian Li. 2020.
\newblock Autoalpha: An efficient hierarchical evolutionary algorithm for mining alpha factors in quantitative investment.
\newblock \emph{arXiv preprint arXiv:2002.08245}.

\bibitem[{Zhao et~al.(2024{\natexlab{a}})Zhao, Jia, Viswanathan, Wu, and Neubig}]{zhao2024self}
Chenyang Zhao, Xueying Jia, Vijay Viswanathan, Tongshuang Wu, and Graham Neubig. 2024{\natexlab{a}}.
\newblock Self-guide: Better task-specific instruction following via self-synthetic finetuning.
\newblock \emph{arXiv preprint arXiv:2407.12874}.

\bibitem[{Zhao et~al.(2024{\natexlab{b}})}]{zhao2024revolutionizing}
H.~Zhao et~al. 2024{\natexlab{b}}.
\newblock \href {http://arxiv.org/abs/2401.11641} {Revolutionizing finance with llms: An overview of applications and insights}.
\newblock Available at: \url{https://arxiv.org/abs/2401.11641} (Accessed: 28 February 2024).

\end{thebibliography}

\appendix
\section{Company List}
\label{sec:comp list}
See Table~\ref{tab:comp table}.
\begin{table}[!h]
\scriptsize
\begin{center}
\begin{tabular}{c |c} 
\hline
\textbf{Sector} & \textbf{Companies}  \\ 
\hline
Information Technology & "AAPL", "AKAM", "AMD" \\
& "ANET", "ANSS", "APH"\\
& "CDNS", "CDW", "CTSH"\\
& "ENPH", "EPAM", "FFIV"\\
& "FSLR", "FTNT", "GEN"\\
& "GLW", "IBM", "INTC"\\
& "IT", "JNPR", "KLAC"\\
& "LRCX", "MCHP", "MPWR" \\
& "MSFT", "MSI", "NOW"\\
& "NXPI", "ON", "PTC"\\
& "QCOM", "ROP", "STX"\\ 
& "SWKS", "TDY", "TEL"\\ 
& "TER", "TRMB", "TXN"\\ 
& "TYL", "VRSN", "WDC", "ZBRA"\\ 
\hline
Health Care &
"ABBV", "ABT","ALGN",\\
& "AMGN", "BAX", "BDX"\\
& "BIO", "BMY", "BSX"\\
& "CAH", "COR", "CRL"\\
& "CTLT", "CVS", "DGX"\\
& "DHR", "DXCM", "EW"\\
& "GILD", "HSIC", "TMO"\\
& "UHS", "VRTX", "VTRS"\\
& "IDXX", "ILMN", "INCY"\\
& "WST", "ZTS", "ISRG", "JNJ"\\
\hline
Energy &
"APA", "COP", "CTRA"\\
& "EOG", "FANG", "HAL"\\
& "HES", "KMI", "MPC"\\
& "MRO", "OKE", "OXY"\\
& "PSX", "PXD", "SLB"\\
& "TRGP", "VLO", "WMB", "XOM"\\
\hline
\end{tabular}
\caption{\label{tab:comp table}Company list of different sectors}
\end{center}
\end{table}

\section{Other Results}
\label{sec:other results}
\subsection{IT Companies with Future One-Month Returns}
\subsubsection{Correlation}
See Figure~\ref{fig:corr2}.
\begin{figure}[!hp]
     \centering
     \begin{subfigure}[!hp]{0.47\textwidth}
         \centering
         \includesvg[width=\textwidth]{it_1m_16-20_old.svg}
         \caption{Correlation of IT companies and future $1$-month returns with existing signals.}
         \label{fig:old signal2}
     \end{subfigure}
     \hfill
     \begin{subfigure}[!hp]{0.47\textwidth}
         \centering
         \includesvg[width=0.75\textwidth]{it_1m_16-20_new.svg}
        \caption{Correlation of IT companies and future $1$-month returns with new signals.}
         \label{fig:new signal2}
     \end{subfigure}
     \hfill
        \caption{Correlation of existing/new signals with returns.}
        \label{fig:corr2}
\end{figure}
\subsubsection{Fama-MacBeth}
\begin{figure}[!hp]
    \centering
    \includesvg[width=0.9\linewidth]{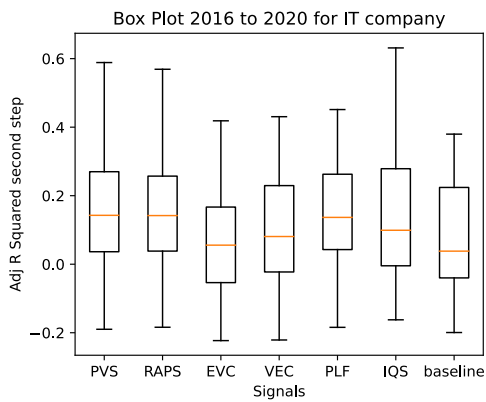}
    \caption{
$R^2_{\text{adj}}$ values of IT companies and future $1$-month returns for Fama-MacBeth step 2. The last boxplot is the baseline without any new signals.  }
    \label{fig:FM step2_2}
\end{figure}

See Figure~\ref{fig:FM step2_2}.

\subsection{Health Care Companies with Future One-Month Returns}
\subsubsection{Correlation}
See Figure~\ref{fig:corr3}.
\begin{figure}[!h]
     \centering
     \begin{subfigure}[!h]{0.47\textwidth}
         \centering
         \includesvg[width=\textwidth]{hc_1m_16-20_old.svg}
         \caption{Correlation of Health Care companies and future $1$-month returns with existing signals.}
         \label{fig:old signal3}
     \end{subfigure}
     \hfill
     \begin{subfigure}[!h]{0.47\textwidth}
         \centering
         \includesvg[width=0.75\textwidth]{hc_1m_16-20_new.svg}
        \caption{Correlation of Health Care companies and future $1$-month returns with new signals.}
         \label{fig:new signal3}
     \end{subfigure}
     \hfill
        \caption{Correlation of existing/new signals with returns.}
        \label{fig:corr3}
\end{figure}
\subsubsection{Fama-MacBeth}
\begin{figure}[!h]
    \centering
    \includesvg[width=0.9\linewidth]{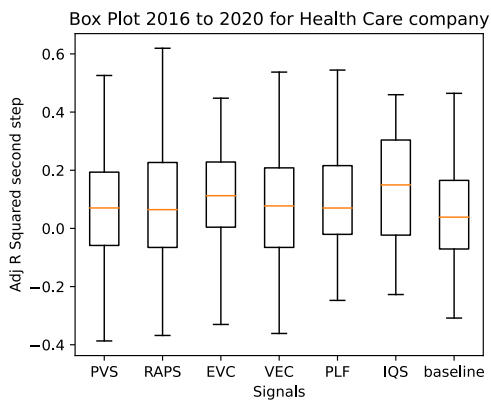}
    \caption{
$R^2_{\text{adj}}$ values of Health Care companies and future $1$-month returns for Fama-MacBeth step 2. The last boxplot is the baseline without any new signals.  }
    \label{fig:FM step2_3}
\end{figure}

See Figure~\ref{fig:FM step2_3}.

\subsection{Health Care Companies with Future Three-Month Returns}
\subsubsection{Correlation}
See Figure~\ref{fig:corr4}.
\begin{figure}[!h]
     \centering
     \begin{subfigure}[!h]{0.47\textwidth}
         \centering
         \includesvg[width=\textwidth]{hc_3m_16-20_old.svg}
         \caption{Correlation of Health Care companies and future $3$-month returns with existing signals.}
         \label{fig:old signal4}
     \end{subfigure}
     \hfill
     \begin{subfigure}[!h]{0.47\textwidth}
         \centering
         \includesvg[width=0.75\textwidth]{hc_3m_16-20_new.svg}
        \caption{Correlation of Health Care companies and future $3$-month returns with new signals.}
         \label{fig:new signal3.5}
     \end{subfigure}
     \hfill
        \caption{Correlation of existing/new signals with returns.}
        \label{fig:corr4}
\end{figure}
\subsubsection{Fama-MacBeth}
\begin{figure}[!h]
    \centering
    \includesvg[width=0.9\linewidth]{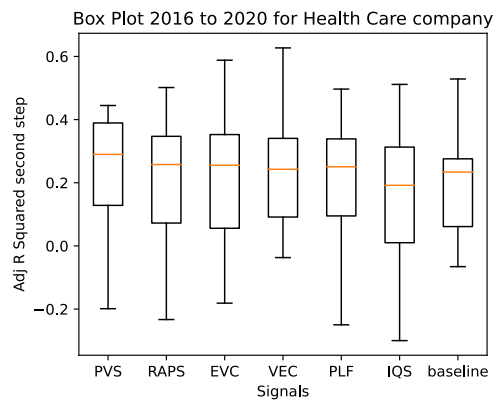}
    \caption{
$R^2_{\text{adj}}$ values of Health Care companies and future $3$-month returns for Fama-MacBeth step 2. The last boxplot is the baseline without any new signals.  }
    \label{fig:FM step2_4}
\end{figure}

See Figure~\ref{fig:FM step2_4}.

\subsection{Energy Companies with Future One-Month Returns}
\subsubsection{Correlation}
See Figure~\ref{fig:corr5}.
\begin{figure}[!h]
     \centering
     \begin{subfigure}[!h]{0.47\textwidth}
         \centering
         \includesvg[width=\textwidth]{en_1m_16-20_old.svg}
         \caption{Correlation of Energy companies and future $1$-month returns with existing signals.}
         \label{fig:old signal5}
     \end{subfigure}
     \hfill
     \begin{subfigure}[!h]{0.47\textwidth}
         \centering
         \includesvg[width=0.75\textwidth]{en_1m_16-20_new.svg}
        \caption{Correlation of Energy companies and future $1$-month returns with new signals.}
         \label{fig:new signal4}
     \end{subfigure}
     \hfill
        \caption{Correlation of existing/new signals with returns.}
        \label{fig:corr5}
\end{figure}
\subsubsection{Fama-MacBeth}
See Figure~\ref{fig:FM step2_5}.
\begin{figure}[!h]
    \centering
    \includesvg[width=0.9\linewidth]{EN_2016_2020_1M_step2.svg}
    \caption{
$R^2_{\text{adj}}$ values of Energy companies and future $1$-month returns for Fama-MacBeth step $2$. The last boxplot is the baseline without any new signals.  }
    \label{fig:FM step2_5}
\end{figure}

\subsection{Energy Companies with Future Three-Month Returns}
\subsubsection{Correlation}
See Figure~\ref{fig:corr6}.
\begin{figure}[!h]
     \centering
     \begin{subfigure}[!h]{0.47\textwidth}
         \centering
         \includesvg[width=\textwidth]{en_3m_16-20_old.svg}
         \caption{Correlation of Energy companies and future $3$-month returns with existing signals.}
         \label{fig:old signal6}
     \end{subfigure}
     \hfill
     \begin{subfigure}[!h]{0.47\textwidth}
         \centering
         \includesvg[width=0.75\textwidth]{en_3m_16-20_new.svg}
        \caption{Correlation of Energy companies and future $3$-month returns with new signals.}
         \label{fig:new signal5}
     \end{subfigure}
     \hfill
        \caption{Correlation of existing/new signals with return.}
        \label{fig:corr6}
\end{figure}
\subsubsection{Fama-MacBeth}
See Figure~\ref{fig:FM step2_6}.
\begin{figure}[!h]
    \centering
    \includesvg[width=0.9\linewidth]{EN_2016_2020_3M_step2.svg}
    \caption{
$R^2_{\text{adj}}$ values of Energy companies and future $3$-month returns for Fama-MacBeth step $2$. The last boxplot is the baseline without any new signals.  }
    \label{fig:FM step2_6}
\end{figure}

\end{document}